\documentstyle[aps,floats,twocolumn]{revtex}

\draft

\begin{document}

\title{\bf On the fourth $P_{11}$ resonance predicted by the constituent
quark model}
\author{L. Theu{\ss}l\thanks{e-mail address: lukas@isn.in2p3.fr}}
\address{
Institut des Sciences Nucl\'eaires, \\
(Unit\'e Mixte de Recherche CNRS-IN2P3, UJF),
  F-38026 Grenoble Cedex, France
}
\author{
R.F. Wagenbrunn\thanks{e-mail address: Robert.Wagenbrunn@kfunigraz.ac.at}}
\address{
Institut f\"ur Theoretische Physik, \\ Universit\"at Graz, 
Universit\"atsplatz  5, A-8010 Graz, Austria 
}

\twocolumn[
    \begin{@twocolumnfalse}
      \maketitle
      \widetext
      \begin{abstract}
We point out a distinguishing difference between constituent quark models based
on one-gluon exchange and one-boson exchange dynamics. 
In the latter one, the $P_{11}$ nucleon resonance with predominantly symmetrical 
spatial 
wave function in the $N=4$ band gets strongly attracted such that it
drops below some states in the $N=2$ band. Calculations of strong decay widths 
are presented in order to establish an identification with experimental states.
Our results are relevant for the interpretation of the fourth $P_{11}$
resonance that was found in the partial wave analysis of the Zagreb group and
recently discussed by Capstick et. al. in the framework of a model based on 
one-gluon exchange.
      \end{abstract}

    \end{@twocolumnfalse}
  ]

{
    \renewcommand{\thefootnote}%
      {\fnsymbol{footnote}}
    \footnotetext[1]{e-mail address: lukas@isn.in2p3.fr}
    \footnotetext[2]{e-mail address: Robert.Wagenbrunn@kfunigraz.ac.at}
  }
\narrowtext

Constituent quark models (CQM's) are useful for interpreting the results of 
partial wave
analyses (PWA) that are usually employed to extract the properties of nucleon
resonances from the experimental data of pion-nucleon 
scattering~\cite{Simula:1999mx}. 
Generally, such models predict four 
$P_{11}$ excited states of the nucleon below or around 2 GeV
which are predominantly part of the $N=2$ band in a harmonic oscillator basis.
The lightest of these states with a totally symmetrical spatial wave function 
is usually attributed to the Roper resonance, the first radial excitation of the
nucleon. Its low mass has presented some problems for conventional CQM's, based
on one-gluon exchange (OGE) dynamics, as these models are not able to describe
the right level ordering of positive and negative parity 
states~\cite{Hogaasen:1983rb}.
The second excitation, with a spatial wave function of mixed symmetry, is attributed to
the $N_{1710}$ resonance, an identification that is confirmed by studies of
decay amplitudes with a specific decay operator~\cite{Koniuk:1980vy}. 
The third state predicted is an
orbital momentum $L=2$ state with quark spin $\frac{3}{2}$ and a 
mixed symmetrical  spatial wave function while the fourth is a $L=1$ state with 
quark spin $\frac{1}{2}$. The latter one has a totally antisymmetrical spatial 
wave function and is therefore naturally expected to be the heaviest of these
resonances with a weak coupling to the $\pi N$ decay channel. 
 
The last two states were not found yet by any partial wave analysis, a fact that
constitutes part of the problem of ``missing states''~\cite{Capstick:1999ng},
i.e. the fact that constituent quark models  generally predict more states than
observed.
In ref.~\cite{Capstick:1999dg} it was argued that the lighter of these missing 
states in the $N=2$ band can be identified with the fourth $P_{11}$
resonance that was discovered by the Zagreb group in their partial wave analysis
based on a multi-channel, multi-resonance, unitary model~\cite{Batinic:1995kr}. 
It was pointed out that the crucial element for
finding this resonance is the fit to the $\pi N \rightarrow \eta N$ data, which
was carried out with great care in ref.~\cite{Batinic:1995kr}. 
Results for partial decay widths in the strong decay
channels $\pi N$, $\eta N$ and $\pi \pi N$, as calculated in 
ref.~\cite{Capstick:1993th} in a relativized version of a model based on OGE
dynamics, were given to confirm the
identification. We redisplay these results in Table~\ref{tab1} together with the
experimental results from ref.~\cite{Batinic:1995kr}. 

\begin{table*}[ht]
\begin{center}
\caption{\label{tab1}
Resonance parameters from different quark models. The first column gives the 
assignment of the Particle Data Group (PDG, ref.~\protect\cite{Groom:2000in}),
OGE is from
ref.~\protect\cite{Capstick:1993th}, GBE are present results and the
experimental data correspond to the four $P_{11}$-resonance solution of 
ref.~\protect\cite{Batinic:1995kr}. For the latter ones we also give the 
uncertainties in the total width that have to be added to the uncertainty in the
partial decay width (the numbers in parentheses). All numbers in MeV.}
\begin{tabular}{lcccccccccccc}
& & \multicolumn{3}{c}{OGE} && \multicolumn{3}{c}{GBE} 
&& \multicolumn{3}{c}{exp.} \\
PDG state && Mass & $\Gamma_{\pi}$ &  $\Gamma_{\eta}$ & &
            Mass & $\Gamma_{\pi}$ &  $\Gamma_{\eta}$ & &
            Mass & $\Gamma_{\pi}$ &  $\Gamma_{\eta}$ \\
\hline
$P_{11}(1440)$ && 1540 & 412 & 0 & & 1443 & 646 & 0  & &
               $1439\pm19$ & $(271\pm18)^{+93}_{-81}$ & $(0\pm0)^{+0}_{-0}$\\
$P_{11}(1710)$ && 1770 & 18  & 67 & & 1770 & 87  & 15 & &
        $1729\pm16$ & $(40\pm40)^{+11}_{-0}$ &$(11\pm11)^{+6}_{-0}$\\
$P_{11}$       && 1880 & 8   & 28 & & 1874 & 70  & 0 & & 
               $1740\pm11$ & $(39\pm39)^{+24}_{-0}$ &$(17\pm13)^{+5}_{-1}$\\
$P_{11}$       && 1975 & 4   & 0 & & 1970 &  5 &  26 & &
               -&-&-\\
$P_{11}(2100)$ && 2065 & 59  & 3 & & 2104 & 4 & 14 & & 
               $2157\pm42$ & $(57\pm17)^{+19}_{-11}$ & $(295\pm18)^{+77}_{-69}$
\end{tabular}
\end{center}
\end{table*}

It can be seen, that the first OGE state that could correspond to the new 
solution found by the Zagreb group at $M\simeq 1740$ MeV is the state 
at $M=1880$ MeV. 
A feature that was not discussed in ref.~\cite{Capstick:1999dg} is the fact that
the model based on OGE predicts this state to decay more strongly into $N \eta$
than into $N \pi$, a situation that seems to be contradicted by the experimental
results (even though one should consider the large error bars in this case). 

The fourth state at $M=1975$ MeV is the totally
antisymmetrical and the heaviest one in the $N=2$ band, as expected. Its weak
coupling to the decay channels was used in particular to explain why it has not
been seen in any partial wave analysis yet. 

The first OGE state in the $N=4$ band 
appears at $M=2065$ MeV and it was assigned to the $P_{11}(N_{2100})$ state 
of the Particle Data Group~\cite{Groom:2000in}, which, however, has the status 
of a 1-star resonance only. 
Its decay properties are very similar to those of the Roper
resonance which indicates the same structure for these two
states. However, as for the relative strength of decay
amplitudes in the $\pi N$ and $\eta N$ channels, the wrong ordering is again
predicted by the OGE as compared to the solution of the Zagreb analysis. 
In fact, just from the decay widths, one would rather guess
that this state corresponds to the new solution of the Zagreb group
at $M\simeq 1740$ MeV, an identification that is hardly possible in this model
because of the large mass of this state in the OGE.

We have calculated the spectrum and wave functions that follow from a model 
based on Goldstone-boson exchange (GBE) 
dynamics~\cite{Glozman:1998ag,Glozman:1998fs}. Here we use a slightly
different parameterization where the tensor components of the
pseudoscalar meson-exchange interaction are included. 
Those tensor forces were dropped in~\cite{Glozman:1998ag,Glozman:1998fs}
and only the spin-spin components of the  pseudoscalar meson-exchange
interaction were taken into account. 

Partial decay widths are then calculated in a microscopic decay model of the 
$^3P_0$ type, which differs from the one used in ref.~\cite{Capstick:1993th}
by two main points: 

\begin{itemize}
\item First we use a $^3P_0$ decay operator in a modified form as
introduced in ref.~\cite{Cano:1996ep}. It has been checked that the 
qualitative features of decay predictions are not influenced by this choice. 
\item Second, and more
important, for all the baryon resonances we use the theoretical masses 
as they follow from the solution of the
Schr\"odinger equation as input in our decay calculation. This would result in
different phase space factors for states whose masses are predicted far off
their experimental values, which is not the case in the GBE for 
any of the resonances
considered here. However it may influence the relative ordering
of  $\pi N$ and $\eta N$ widths due to the 
``structure dependence''~\cite{Koniuk:1980vy} of
the $P_{11}$ resonances\footnote{The decay amplitude for  $P_{11}$ 
resonances
has a zero as a function of the (off-shell) meson momentum $q$. Depending on
whether one chooses $q$ to be consistent with the theoretical or with the
experimental resonance mass, one may approach this zero or move away from it.}.
\end{itemize}

First results in this approach were presented in ref.~\cite{Theussl:2000sj},
where also a more thorough discussion regarding constituent quark- and 
decay models may be found. The masses of
the lightest resonances predicted by the GBE in the $P_{11}$ channel are given 
in Table~\ref{tab1}
and it can be seen that, apart from the Roper resonance, which is better
described by the GBE, they correspond almost exactly to the masses of the OGE
model. The decay properties are rather different however. 

First we note that for the $N_{1710}$ resonance the GBE model predicts 
$\Gamma_{\eta}<\Gamma_{\pi}$, contrary to the OGE model but in qualitative 
agreement with the experimental result of the Zagreb analysis. 
A similar difference appears
for the next $P_{11}$ resonance, where again the GBE reproduces roughly the
experimental pattern of decay amplitudes, whereas the OGE predicts the opposite
ordering.

In order to explain these qualitative differences we display in Table~\ref{tab2} 
the $LS$ - components of the GBE wave functions in question.

\begin{table}[ht]
\begin{center}
\caption{\label{tab2}
Probabilities (in \%) of all possible $LS$ - components of 
$N^*$ wave functions for $P_{11}$ resonances 
from the GBE model with tensor force.}
\begin{tabular}{ccccccccc}
&&&&\multicolumn{2}{c}{$S=1/2$} && \multicolumn{2}{c}{$S=3/2$} \\ 
$[N ^{J^{\pi}}]_n$ & $ N^* $ & Mass & & L=0 & L=1 & & L=1 & L=2  \\ 
\hline
$[N ^{\frac{1}{2}^+}]_1$ & $N_{939}$  & 939  && 98.7 &      &&     & 1.3  \\
$[N ^{\frac{1}{2}^+}]_2$ & $N_{1440}$ & 1443 && 98.9 &      &&     & 1.1  \\
$[N ^{\frac{1}{2}^+}]_3$ & $N_{1710}$ & 1770 && 96.9 &      &&     & 3.1  \\
$[N ^{\frac{1}{2}^+}]_4$ &            & 1874 && 98.5 & 0.1  &&     & 1.4  \\
$[N ^{\frac{1}{2}^+}]_5$ &            & 1970 && 2.9  & 43.3 && 0.1 & 53.7 \\
$[N ^{\frac{1}{2}^+}]_6$ & $N_{2100}$ & 2104 && 1.8  & 54.7 && 0.8 & 42.7 \\
\end{tabular}
\end{center}
\end{table}

It is seen immediately, that the GBE state at $M=1874$ MeV 
cannot be identified with any of the remaining states in the $N=2$ band, 
since it is an almost pure $L=0$ state. It rather
corresponds to the predominantly symmetrical solution in the $N=4$ band,
i.e. the second radial excitation of the nucleon after the Roper resonance.
The properties of the Goldstone-boson exchange dynamics lead to a 
strong attraction in symmetrical components of the wave function, a property that
explained already the low mass of the Roper resonance in this model.

We see also from Table~\ref{tab1} that the decay properties correspond almost
exactly to those of the $N=4$ state in the OGE model. In particular, we
predict the fourth $P_{11}$ resonance to decay more strongly into $N \pi$
than into $N \eta$, as it was found for the new resonance in the
Zagreb analysis.

The last two GBE states given in Table~\ref{tab1} are found to be almost 
50\% - 
mixtures of the $L=2$, $S=\frac{3}{2}$ and 
$L=1$, $S=\frac{1}{2}$ components. These ones receive a repulsive contribution
from the GBE interaction, because they contain quark pairs with antisymmetrical
spin-flavor wave functions. This enhances the effect of level inversion with the
symmetrical state in the $N=4$ band.

Due to the different symmetry structure it is also expected that
they should have smaller branching fractions into the $N\pi$ channel 
than the lower $P_{11}$
resonances, a feature that is shared with the OGE predictions for the
corresponding states. A qualitative difference
between the two models again occurs when comparing the $N\eta$ with the
$N\pi$ widths: both these resonances give $\Gamma_{\eta}>\Gamma_{\pi}$ in the
GBE, a relation that is also satisfied by the state at $\sim 2157$ MeV found in
the Zagreb analysis, but contradicted by the predictions of the OGE. This
feature is a consequence of the strong mixing of these two states, 
as evidenced in Table~\ref{tab2}. In fact, the main
contribution to the decay widths for both of them comes from the $L=2$, 
$S=\frac{3}{2}$ component which explains their similar decay properties as
opposed to the OGE model.

We would finally like to stress that the inversion of states discussed above 
is also found in
different parameterizations of the GBE interaction. We have checked in
particular that in the models of ref.~\cite{Glozman:1998ag,Glozman:1998fs} as
well as ref.~\cite{Wagenbrunn:2000ue,Wagenbrunn:2000sg}, it is always the mainly
symmetrical state that follows the $N_{1710}$ resonance. A good description of
the Roper resonance has also been obtained in a model including a three-body
force~\cite{Desplanques:1992rf}. 
Since this one acts mainly on the nucleon and its
radial excitations while producing essentially no effect on states with mixed
symmetry, one may expect a similar inversion in the upper part of the spectrum
as in the GBE model. Unfortunately, calculations of decay widths as presented 
in ref.~\cite{Cano:1996ep} do not include the states of interest here.

In addition to $\pi N$ and $\eta N$ widths, also results for the $\pi\pi N$
channel were given in ref.~\cite{Capstick:1993th} and compared to the results of
the Zagreb group. We did not repeat these
calculations in the GBE model because of some theoretical ambiguities associated
with the determination of a decay operator for 
quasi-two-body decays~\cite{Capstick:1994kb}. Furthermore the branching
fractions extracted from the PWA depend sensitively on the input data of $\pi N
\rightarrow \pi\pi N$ reactions, but the inelastic data in channels like
$\pi\Delta$, $\rho N$, etc., were not explicitly included in the analysis of
the Zagreb group~\cite{Batinic:1995kr}. 

In all the considerations presented above, 
one has to bear in mind the rather large
uncertainties of experimental data which prevents one from drawing definite
conclusions about the quality of predictions from any CQM for the moment. 
However, the inversion of states in the GBE model is a
unique and distinguishing feature leading to qualitatively different
predictions of decay properties. This would permit a definite discrimination
of different models, once the experimental data are determined with 
sufficient accuracy.

In summary, we have pointed out that the states  that are 
predicted by models based on OGE and GBE dynamics at almost the same mass of
$M\simeq 1880$ MeV do not correspond to the same SU(6) $\otimes$ O(3) state in a
harmonic oscillator basis. As a consequence they show very different decay 
properties. In particular, predictions for decay widths using wave functions
stemming from GBE dynamics are in qualitative agreement with the recent PWA of
ref.~\cite{Batinic:1995kr}. Differences with predictions from the OGE model may
allow a discrimination between the two models, which is not possible at the
moment due to rather large experimental uncertainties.
Clearly, a more precise determination of the resonance parameters discussed 
in this paper is needed to definitely settle the issue.

\bigskip
{\bf Acknowledgements}
We are grateful to  Fl. Stancu and B. Desplanques for useful discussions and to
the ECT$^*$, Trento, for the support and hospitality during the time when this 
work was begun.
This work was supported by the Scientific-Technical Agreement 
'Amad\'ee' between Austria and France under contract number II.9. 


\end{document}